%% file: comsnets2020.tex
\documentclass[conference]{IEEEtran}
\IEEEoverridecommandlockouts
\usepackage{cite}
\usepackage{amsmath,amssymb,amsfonts}
\usepackage{algorithmic}
\usepackage{graphicx}
\usepackage{textcomp}
\usepackage{xcolor}
\usepackage{subcaption}
\usepackage{listings}
\usepackage[draft, bookmarks=false]{hyperref}
\def\BibTeX{{\rm B\kern-.05em{\sc i\kern-.025em b}\kern-.08em
    T\kern-.1667em\lower.7ex\hbox{E}\kern-.125emX}}
\begin{document}

\title{
Improving the Performance of Deep Learning\\ for Wireless Localization
%Hyperparameter Tuning and Data Augmentation for Deep Learning in Wireless Localization
%Deep Learning for Indoor Wireless Localization: Issues and Resolutions
}

\author{\IEEEauthorblockN{
Ramdoot Pydipaty\IEEEauthorrefmark{1},
Johnu George\IEEEauthorrefmark{2},
Krishna Selvaraju\IEEEauthorrefmark{3}
and Amit Saha\IEEEauthorrefmark{4}}
\IEEEauthorblockA{
%\IEEEauthorrefmark{1}\IEEEauthorrefmark{2}\IEEEauthorrefmark{4}Cisco Systems, Bangalore, India, \IEEEauthorrefmark{3} HCL Technologies, India\\
Cisco Systems, Bangalore, India\\
Email: \IEEEauthorrefmark{1}rapydipa@cisco.com,
\IEEEauthorrefmark{2}johnugeo@cisco.com,
\IEEEauthorrefmark{3}duselvar@cisco.com,
\IEEEauthorrefmark{4}amisaha@cisco.com\\
}
%\thanks{\IEEEauthorrefmark{3}Author was working for Cisco Systems, India for this project.}
}

\maketitle

\begin{abstract}
Indoor localization systems are most commonly based on Received Signal Strength
Indicator (RSSI) measurements of either WiFi or Bluetooth-Low-Energy (BLE) beacons. In such
systems, the two most common techniques are trilateration and fingerprinting, with the latter
providing higher accuracy. In the fingerprinting technique, Deep Learning (DL) algorithms 
are often used to predict the location of the receiver based on the RSSI measurements
of multiple beacons received at the receiver.
%Off late, Deep Neural Networks (DNNs) are being explored to improve the predictions.

%DNNs are however notorious to tune since there are several hyper-parameters to be optimized.
%This makes such solutions difficult to port to newer wireless setups since the tuning is 
%specific to each setup.

%In this paper, we present the design and analysis of a
In this paper, we address two practical issues with applying Deep Learning
to wireless localization --- transfer of solution from one wireless environment
to another \emph{and} 
small size of labelled data set. First, we apply automatic
hyperparameter optimization to a deep neural network (DNN) system for indoor
wireless localization, which makes the system easy to port to new wireless
environments.
%We evaluate it using a freely available BLE RSSI dataset and show
%that our proposal allows automatic tuning of the DNN system. This makes the system
%easy to port to new wireless environments.
%Second, we show how beacons can be selectively remove
%beacons from the data set to figure out the
%impact of each beacon on localization performance. Finally, we show how to augment
Second, we show how to augment
a typically small labelled data set using the unlabelled data set.
We observed improved performance
in DL by applying the two techniques. Additionally, all relevant
code has been made freely available.

%Additionally, for the benefit of
%the community, we have made our code and system open-source.
\end{abstract}


\input{tex/introduction.tex}
\input{tex/hyperopt.tex}

\input{tex/dataset.tex}
\input{tex/implementation.tex}
\input{tex/dnn_and_cnn.tex}

\input{tex/augmentation.tex}
\input{tex/conclusion.tex}

%\section*{Acknowledgment}

\bibliographystyle{./bibliography/IEEEtran}
%\bibliography{./bibliography/IEEEabrv,./bibliography/IEEEexample}
\bibliography{./bibliography/IEEEabrv,./bibliography/mypaper.bib}

\appendix
\input{tex/appendix.tex}

\end{document}

%% file: tex/introduction.tex
\section{Introduction}
\label{introduction}
Systems like GPS have made outdoor localization very prevalent. However, in an indoor setting,
GPS does not work and other approaches have to be taken to perform localization. Indoor 
localization has a lot of uses such as device and personnel tracking, targeted advertising,
and broadcasting localized information. 

One of the most common ways of performing indoor localization is to use the reception
of wireless signals at any wireless receiver, such a laptop or mobile handset.
In particular, the Received Signal Strength Indicator (RSSI) at any receiver has a
correlation with the location of the source of the signal. 
The two most common ways of using RSSI measurements to predict the location of the receiver
are {\it trilateration} (sometimes called triangulation) and {\it fingerprinting}.
Trilateration~\cite{liu:survey2007} is the process of
using the estimated distances from multiple sources to 
triangulate a most probable location of the receiver. The earliest works in localization
have used this approach~\cite{bahl:infocom2000}. Obviously, the radio frequency
propagation of the environment has a great impact on the accuracy of this method.
For example, effects like multi-path can cause a lot of noise in the received signal. 

Fingerprinting~\cite{journal:fingerprint}
on the other hand is somewhat more resilient to the idiosyncrasies of the 
environment because it relies on prior measurements being collected for multiple known
locations in a specific
environment -- one fingerprint for each measured location, so to say. This takes care of 
any vagaries of the environment affecting the path loss propagation model. Of course any
drastic change in the environment will make the measurements less relevant.
Once the measurements are collected, the system then predicts
the location of the receiver by comparing the RSSI measurements with the fingerprints for
different locations and selecting the one with the most similar fingerprint as the 
most probable location of the receiver. 
%As specified Tian et al.~\cite{Tian2013},
%fingerprinting approaches can be further divided into deterministic, probabilistic, 
%and ML-based, with only the ML-based approach being relevant to this paper.

Due to the prevalence of IEEE 802.11-WiFi deployments,
the initial fingerprinting based systems used
the RSSI measurements of WiFi beacons. However, 
Bluetooth Low Energy~(BLE) is fast becoming an alternate to WiFi access points since BLE 
transmitters consume much less energy compared to WiFi access points. This allows
BLE transmitters to run on battery for months and even years and hence can be installed
without any worry of access to power. Additionally, the BLE approach
seems to be able to provide
more fine grained localization than WiFi based solutions,
as reported by Zhao et al.~\cite{zhao:bleVSwifi2014, faragher:ble, hindawi:ble}.
The reasons behind why
BLE outperforms WiFi for localization is outside the scope of this paper.

\subsection{Issues with Deep Learning in Wireless Localization}
There are several ML algorithms that are used in the fingerprinting approach.
Some of these 
are $k$-nearest neighbors, support vector machines,
probabilistic methods (e.g., Naive 
Bayes), and {\it deep neural networks (DNN)} --
a good survey is presented by Liu et al.~\cite{liu:survey2007} and
Bozkurt et al.~\cite{bozkurt:comparision}.
However, using ML, specifically DNNs, for this 
has several issues and in this paper, we address the following two major issues.

First, each layout/floor map would need a different model
or one may use the same model architecture and tune it for each layout.
%So, \emph{automatic hyperparameter tuning is important for finding a better model}
%for any given layout.
Even without any change in the layout, since wireless environment is dynamic
due to multi-path effect, noise, etc., 
\emph{frequent tuning of the models is a necessity}.
Automatic tuning of the model gives us a quick,
scalable way of tuning, thus improving models.
A typical problem in using DNNs (used in any problem domain
and not just RSSI fingerprinting)
is that these neural networks usually have quite a few
hyperparameters (knobs), such as {\it learning rate}, {\it number of hidden nodes}, and 
{\it batch size},
that have to be set before the training process. These knobs
generally have a large impact on the metrics of the neural network, such as training time
and  accuracy and hence tuning these knobs is essential to improve the performance of the
neural network. Consequently, figuring out the optimum values for these knobs
(hyperparameters) is non-trivial. 
So, \emph{automatic hyperparameter tuning (HP-tuning) is important for finding a better model}
for any given layout.
In this paper, we show how hyperparameter 
optimization can be applied to a DNN (that would otherwise have to be manually tuned)
to perform RSSI fingerprint matching.

%Secondly, given any set of measurements, an obvious need is to {\it rationalize the number
%of the beacons} thus justifying whether the number of beacons used for a particular 
%environment is just right and if there are beacons which are more important than others.

%Finally, a typical problem in data collection for wireless localization is that {\it 
Second, a typical problem in data collection for wireless localization is that {\it 
number of labelled data points
is largely outnumbered by the number of unlabelled data points.} Due to this lack of 
labelled data, the ML algorithms trained on the data are difficult to generalize, thus making
%the ensuing models less portable.
the learning less transferable to other wireless environments. In this work, we 
present augmentation techniques for augmenting the unlabelled samples.

\subsection{Related Work}
\label{related}

As mentioned in Section~\ref{introduction}, there have been several
pieces of work in using ML in the area of fingerprint based
indoor localization~\cite{liu:survey2007}. There has also been work on
using principal component analysis (PCA) to reduce computation cost of ML-based
WiFi indoor localization~\cite{salamah:localization_ml}.
Here we explain some deep learning
based work that are relevant to our work. 
Zhang et al.~\cite{Zhang:2016:DNN:2936025.2936204} addressed important issues with
deep neural networks for
both indoor and outdoor wireless localization. However, they did not analyze the 
use of HP-tuning or
%rationalization of number of beacons,
augmentation of the data set. To the best of our knowledge we could not find any prior work
related to the application of hyperparameter tuning (HP-tuning)
to DNNs for wireless localization.
%Recently, Tu et al.~\cite{tu:kdd2019}, applied
%HP-tuning to the problem of network embedding but that is not relevant for our work.
Similarly, there are augmentation techniques available for image related machine learning 
models~\cite{Shorten2019}
but we could not find anything related to augmentation of wireless RSSI samples.

%For beacon rationalization, we use a dropout technique. Dropout was first
%introduced by Hinton et al.~\cite{mypaper:dropout} and there are several ways 
%of doing dropout and it is now a standard technique for regularizing ML networks.
%However, we could not find the use of dropout for studying the impact of removing
%beacons on wireless localization using DNNs.

One of the problems in working with existing DNN models for wireless localization is that
the code for the models are often not publicly available making it difficult to analyze
and extend the work. In this respect, Kaggle~\cite{mypaper:kaggle} is a widely used
source of publicly available models
making it easy to analyze the code, fix any issues, and 
contribute back extensions or corrections to the models; we chose this path
to get a DNN from Kaggle\cite{mypaper:kaggle-dnn}. We also introduced a CNN model of our
own. We used these as the base code to perform HP-tuning.

\subsection{Our Contributions}
This paper addresses the issue mentioned above and our main contributions can 
be summarized as follows:
\begin{itemize}
\item We present the design and analysis of applying HP-tuning
to a deep neural network (DNN) system for indoor wireless localization.
%\item We show how we can selectively remove beacons from data set to figure out which 
%beacons are the most important and whether all beacons are required or not.
\item We introduce a convolutional neural network (CNN) model for wireless localization and
apply HP-tuning to it as well.
\item We present an augmentation technique to augment the labelled data sets with data points
from the unlabelled data set.
\item We have also made all relevant code public so that our work can be easily 
verified and extended.
%\item We present an open source framework for performing hyperparameter optimization
%of deep learning models for wireless localization.
\end{itemize}

The rest of this paper is arranged as follows.
%Section~\ref{related} presents the prior
%art in the use of machine learning (ML) for fingerprinting based localization.
Section~\ref{hyperopt} gives an overview of hyperparameter optimization. We explain
the data set used in this paper in Section~\ref{dataset}. In Section~\ref{implementation},
we give an overview of our
implementation setup used to run all our experiments.
Section~\ref{dnn_and_cnn} explains
the DNN and CNN models and the impact of applying HP-tuning to them.
We present how the labelled data
set can be augmented using data
points from the unlabelled data set in Section~\ref{augmentation}.
%and in Section~\ref{rationalize} we present our idea for beacon rationalization.
Finally, we conclude in Section~\ref{conclusion}.

%% file: tex/hyperopt.tex
\section{Background: Hyperparameter Optimization}
\label{hyperopt}

In machine learning, a hyperparameter is a parameter that must be fixed before the actual
training process. Consequently, hyper-parameters (e.g., learning rate, batch size, 
and number of hidden nodes) 
cannot be learnt during the learning process, 
unlike the value of parameters (e.g., weights) that are learned during the training process.
Hyperparameters can impact both the quality of the model generated by the training process as
well as the time and memory requirements of the algorithm~\cite{Goodfellow-et-al-2016}.
Thus, hyperparameters have to be tuned to get the optimal setting for a given problem. This
tuning can be done either manually or automatically. Manual tuning might suffice if the 
problem setup is not expected to change, at least not frequently. In case the problem setup 
is expected to change frequently, an automatic hyperparameter tuning (HP-tuning)
mechanism is required to make the system practical.

%\subsection{Hyperparameter Optimization Search Algorithms}
%\label{algos}
There are several common algorithms that are used in automatic hyperparameter tuning. Some
of the most common are: 
\begin{itemize}
    \item \emph{Grid search}: Given a range of possible values for each of the hyperparameters, 
    this exhaustive
    algorithm runs experiments with all hyperparameter values in the Cartesian product 
    of the set of values for each individual hyperparameter. This technique works best 
    when there are few hyperparameters ($<=3$).
    \item \emph{Random search}: This improves upon the grid search method by searching the
    range of each hyperparameter randomly rather than exhaustively. This leads to faster search
    times and hence works better than grid search, especially when the number of
    hyperparameters is large.
    \item \emph{Bayesian optimization}: In contrast to grid or random search, Bayesian
    optimization uses a probability model of the objective function and uses that to select
    the most promising hyperparameter values~\cite{Bergstra:2013:MSM:3042817.3042832}.
    \emph{We use this search technique in this paper}.
\end{itemize}

\begin{figure}[t]
    \centering
    \includegraphics[width=\linewidth]{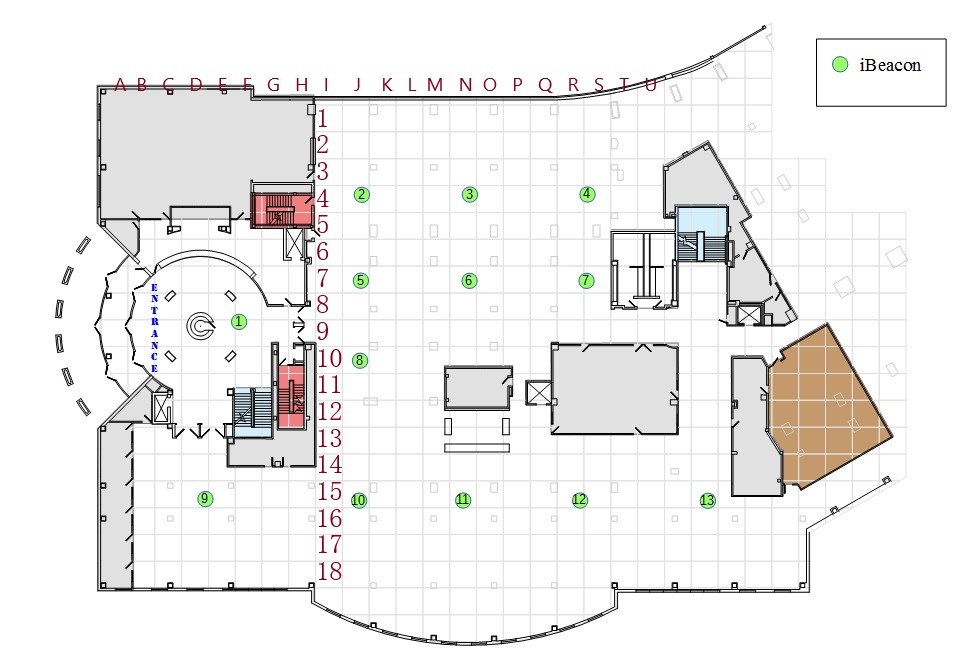}
    \caption{The experimental setup with 13
    iBeacons in a $200 ft.$ X $180 ft.$ 
    library~\cite{UCI:ble-dataset, mohammadi:ble-dataset}.}
    \label{fig:setup}
\end{figure}

%% file: tex/dataset.tex
\section{Data Set}
\label{dataset}
Our evaluation is based on a real world data set that is available at the 
Machine Learning Repository, made public by the University of
California, Irvine~\cite{UCI:ble-dataset}. This data set was first presented by
Mohammadi et al.~\cite{mohammadi:ble-dataset} and much of the description of the data set
presented here has been taken from the original paper.

The data set is from a real-world deployment of 13 of Apple's iBeacons, which 
use the Bluetooth Low Energy~(BLE) standard of Bluetooth 4.0, in a campus library of
dimension 200 ft. X 180 ft. These beacons are installed on the ceiling of the first floor
of the Waldo Library in Western Michigan University, USA. On an average, each of the 13
iBeacons are at a distance of around 30-40 ft. from nearby iBeacons. The entire floor
space was divided in 10 ft. X 10 ft. block and the RSSI
measurements were taken at several 
locations, with each location being manually captured.
Each location is usually covered by multiple beacons and all those measurements
were captured.
The entire labelled data set contains 1136 data points used for training and
284 data points for testing,
which is a typical 80:20 split of the total 1420 labelled data points.
In addition to the labelled samples, there are 5191 unlabelled samples as well.
As mentioned in Section~\ref{introduction}, this is a typical problem in sample
collection and we will address this issue in Section~\ref{augmentation}.

\begin{table}[t]
\begin{center}
    \resizebox{\columnwidth}{!}{%
\begin{tabular}{||l | c ||}
    \hline
    {\bf Setup/Parameter} & {\bf Value}\\
    \hline\hline
    Hyperparameter Tuner & Katib\\
    \hline
    ML framework & Keras\\
    \hline
    Epochs & 100\\
    \hline
    Batch Size & 100\\
    \hline
    Activation Function & ReLU\\
    \hline
    Loss Function & Root mean squared error (rmse)\\
    \hline
    Optimizer & Adam, SGD\\
    \hline
    Number of parameters & DNN: 6102, CNN: 5438\\
    \hline
    Training data points & 1136\\
    \hline
    Test data points & 284\\
    \hline
    \hline
\end{tabular}
}
\caption{Training Setup and Parameters. A single forward and backward pass
of the entire data set through the neural network is defined as an {\it epoch}.
Since entire data set might be too large to train in one shot,
the data set is divided into {\it batch size} chunks.}
\label{tab:params}
\end{center}
\end{table}

%% file: tex/implementation.tex
\section{Implementation Details}
\label{implementation}
This section describes the implementation details for the rest of the paper.
We provisioned a Kubernetes~\cite{mypaper:k8s} cluster, 
%and deployed Kubeflow~\cite{mypaper:kubeflow} and Katib (Appendix~\ref{katib}),
%and deployed Kubeflow (Appendix~\ref{kubeflow}) and Katib (Appendix~\ref{katib}),
and deployed Katib (Appendix~\ref{katib}),
the open source hyperparameter tuner, on Kubeflow (Appendix~\ref{kubeflow}), the ML
toolkit for Kubernetes.
%following https://www.kubeflow.org/docs/components/hyperparameter-tuning/hyperparameter/.  
We implemented all the ML models in Keras~\cite{mypaper:keras}.
We packaged each Keras model,
which takes hyperparameters as input and emits loss metric as the output,
into a Docker~\cite{mypaper:docker} image. 
These images are used in Katib user interface
\emph{trial specification} to trigger the \emph{trials}.
The range of hyperparameters is set accordingly 
for each optimizer in the user interface,
e.g., for Adam optimizer, \emph{learning rate} -- [0.001, 0.002],
\emph{beta1} -- [0.88, 0.93].
The objective metric to be minimized is set to Euclidean/l2 loss,
with a goal of 1.2, specified in the user interface.
Also, we limited the maximum trials for each tuning experiment to 15 and chose
the Bayesian optimization algorithm. All code relevant to this paper has been
made public at
{\it \url{https://github.com/CiscoAI/deep-learning-wireless-localization}}.

%% file: tex/dnn_and_cnn.tex
\section{DNN and CNN for Wireless Localization}
\label{dnn_and_cnn}
As mentioned earlier, deep neural networks have been used since quite sometime to
address wireless localization, especially using the fingerprinting technique.

\subsection{DNN}
\label{dnn}

\begin{figure}[tbp]
    \centering
    \includegraphics[width=0.9\linewidth]{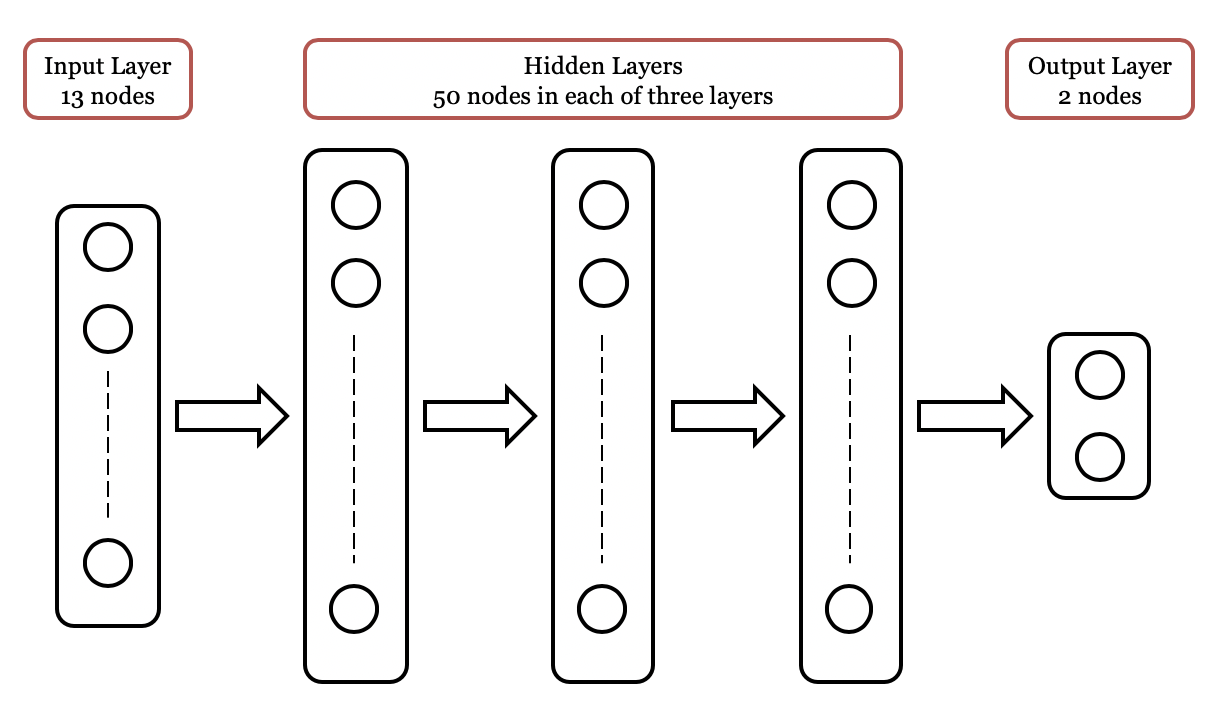}
    \caption{DNN Architecture. 
    $[${\bf Input layer}: 13 nodes, 1 each for each iBeacon; 
    {\bf Hidden layers}: 3 layers, each with 50 nodes; {\bf Output layer}: 2 nodes,
    representing (x,y) coordinates of predicted location; 
    {\bf Activation function}: ReLU$]$}
    \label{fig:dnn}
\end{figure}

\begin{figure*}[tbp]
     \begin{subfigure}{0.5\textwidth}
        \centering
        \includegraphics[width=\linewidth]{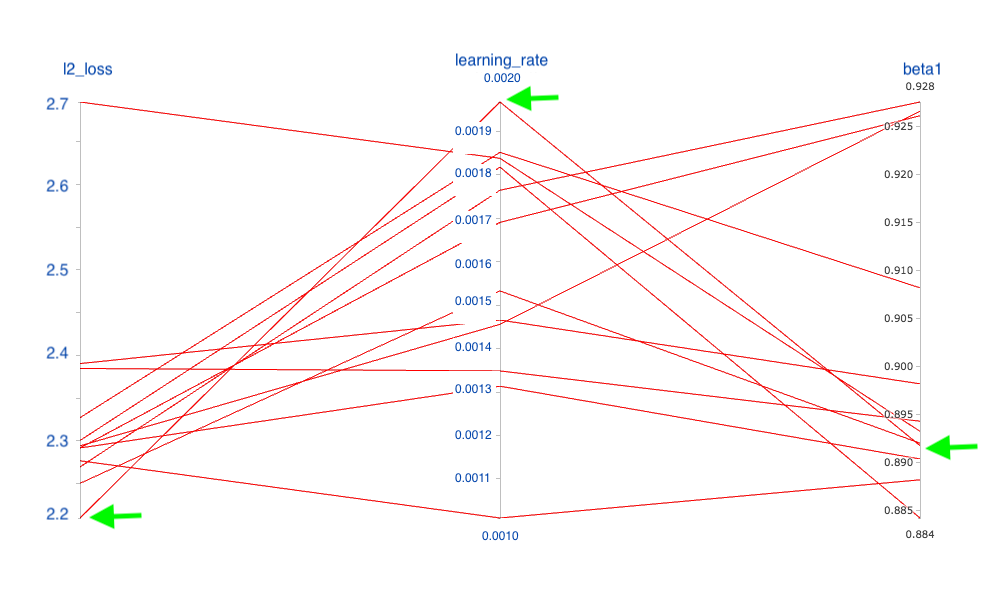}
        \caption{Adam Optimizer DNN.}
        \label{fig:adam_dnn}
    \end{subfigure}
    \begin{subfigure}{0.5\textwidth}
        \centering
        \includegraphics[width=\linewidth]{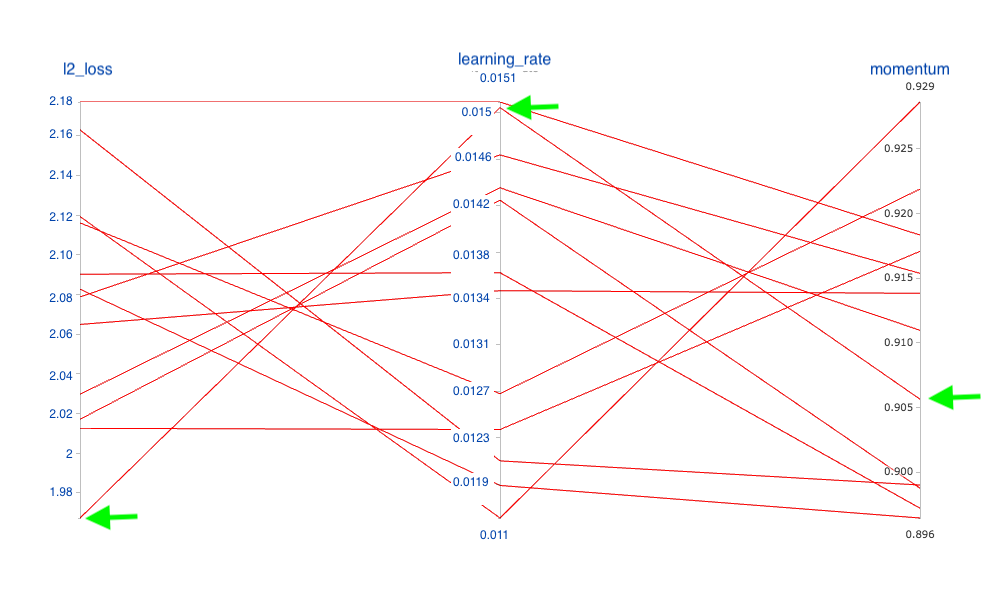}
        \caption{SGD Optmizer in DNN.}
        \label{fig:sgd_dnn}
    \end{subfigure}
    \caption{Hyperparameter optimization using Bayesian optimization applied to DNN.
    The arrows mark the optimal values.
    }
\label{fig:dnn-bayesian-opt}
\end{figure*}

\begin{figure}[tbp]
    \centering
    \includegraphics[width=1.0\linewidth]{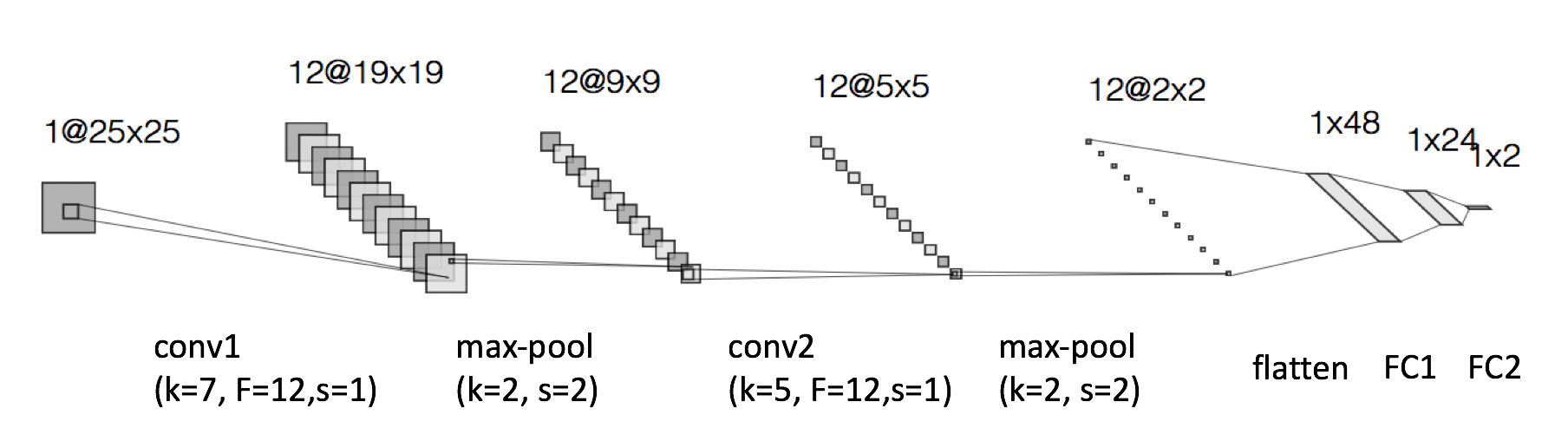}
    \caption{The CNN architecture.
    $[${\bf Input layer}: [25, 25, 1] grayscale image;
    {\bf 2 Convolution layers}: number of filters respectively [12, 12];
    {\bf Filter sizes}: [7, 7], [5, 5] with valid padding;
    {\bf 1 Dense layer}: 24 nodes;
    {\bf Output layer}: 2 representing $x$ and $y$ coordinates;
    {\bf Activation function}: ReLU$]$}
    \label{fig:cnn}
\end{figure}

A deep neural network (DNN) is any neural network that has layers
in addition to the input and output 
layers; these additional layers are known as hidden layers.
The number of hidden layers can vary from
a few (2--3) to thousands. For our study we used an existing DNN
model~\cite{mypaper:kaggle-dnn}, shown
here in Figure~\ref{fig:dnn}. The DNN takes in each of the 13 RSSI measurements 
and tries to predict the \emph{$<$x,y$>$}
coordinates of the location where the measurement was made.
The data set described in Section~\ref{dataset}
was used with a 80:20 (1136:284) 
train:test split. The training parameters are specified in Table~\ref{tab:params}.

\subsection{DNN with Hyperparameter Optimization}
\label{dnn_hp}

\begin{table*}[t]
\begin{center}
    \resizebox{\textwidth}{!}{%
\begin{tabular}{||l | l || p{0.5in} | p{0.5in} | p{0.5in} | p{0.5in} | p{0.65in} | p{0.65in} | p{0.65in} | p{0.85in} ||}
    \hline
    {\bf Sl. No} & {\bf Setup} & {\bf default learning rate} & {\bf tuned learning rate}
    & {\bf default beta1} & {\bf tuned beta1} & {\bf default momentum}
    & {\bf tuned momentum} & {\bf loss metric with default} 
    & {\bf loss metric with tuned parameters} \\
    \hline\hline
    1. & DNN + Adam & 0.001 & 0.00197 & 0.9 & 0.89177 & --- & --- & 2.32 & 2.2\\
    \hline
    2. & DNN + SGD & 0.01 & 0.015 & --- & --- & 0.9 & 0.905 & 2.10 & 1.97\\
    \hline
    3. & CNN + Adam & 0.001 & 0.00177 & 0.9 & 0.885 & --- & --- & 2.4 & 2.11\\
    \hline
    4. & CNN + SGD & 0.01 & 0.01375 & --- & --- & 0.9 & 0.885 & 2.36 & 2.10\\
    \hline
    \hline
\end{tabular}
}
\caption{Hyperparameter values and loss metrics. Loss metrics are in units of 
grid length, i.e., 2.2 means 2.2 x 10ft = 22 ft.}
\label{tab:hyperparams}
\end{center}
\end{table*}

As explained earlier, the performance of a model is affected by the choice of the
hyperparameters and figuring out the optimal value for of the hyperparameters is 
non-trivial. This issue is magnified if the underlying wireless environment changes.
Consequently, we apply hyperparameter tuning (HP-tuning)
to the DNN and check if it provides any
advantage.
The result of the experiment are shown in Figure~\ref{fig:dnn-bayesian-opt}. The 
HP-tuning was tried for both the Adam optimizer~\cite{mypaper:adam}
and the Stochastic Gradient Descent (SGD) optimizer~\cite{mypaper:sgd}
and the results are shown respectively
in Figure~\ref{fig:adam_dnn} and Figure~\ref{fig:sgd_dnn}. For the Adam optimizer, 
the hyperparameters are the \emph{learning rate}, \emph{beta1},
and \emph{beta2} and for the SGD the hyperparameters are \emph{learning rate} and
\emph{momentum}. In order to make both graphs comparable, we just used \emph{beta1}
and left out \emph{beta2}. The figures show how the different values of the 
hyperparameters impact the metric, the \emph{l2 loss}. 
From Figure~\ref{fig:dnn-bayesian-opt} the following can be observed (as consolidated
in Table~\ref{tab:hyperparams}):
\begin{itemize}
    \item The best combination for the Adam optimizer (Figure~\ref{fig:adam_dnn})
    is a \emph{learning rate} value of 0.002 and \emph{beta1} value of 0.893.
    This gives an \emph{l2 loss} of 22 feet.
    \item The best combination for the SGD optimizer (Figure~\ref{fig:sgd_dnn})
    is an \emph{learning rate} value of 0.015 and \emph{momentum} value of 0.905.
    This gives and \emph{l2 loss} of 19.7 feet.
    \item Even a small change in the value of \emph{learning rate} can
    significantly impact the metric.
    \item The \emph{l2 loss} varies from 22.0ft to 27.0ft for 
    the Adam optimizer (Figure~\ref{fig:adam_dnn}) 
    and from 19.6ft to 21.8ft for the SGD optimizer
    (Figure~\ref{fig:sgd_dnn}). This means
    that the SGD optimizer is a better optimizer for this problem compared to the 
    Adam optimizer.
\end{itemize}

Apart from \emph{learning rate}, other hyperparameters such as \emph{batch size},
\emph{number of layers}, and \emph{layer sizes} can also be tuned using the tuning
framework, thus allowing an optimal search of the entire search space to find the 
best set of hyperparameters. However, we did not perform this exhaustive
search for this paper.

\subsection{CNN}
\label{cnn}
A DNN has a
major disadvantage that the model is tied to the number of beacons -- if
there is a change in the number of beacons, the model has to be reconstructed since
the same model might not work for scenarios with a large difference in the number of 
beacons. In order to handle this, a special type of DNN, called a Convolutional
Neural Network (CNN) can be used. CNNs are neural networks that use
{\it convolutions}, a special kind of linear operation, as opposed to the typical
matrix multiplication, in at least one of the layers of the neural network. CNNs are 
most useful when the data being processed has a grid like topology such as, most
commonly, image data, but also time series data~\cite{Goodfellow-et-al-2016}.

\begin{figure*}[t]
     \begin{subfigure}{0.5\textwidth}
        \centering
        \includegraphics[width=\linewidth]{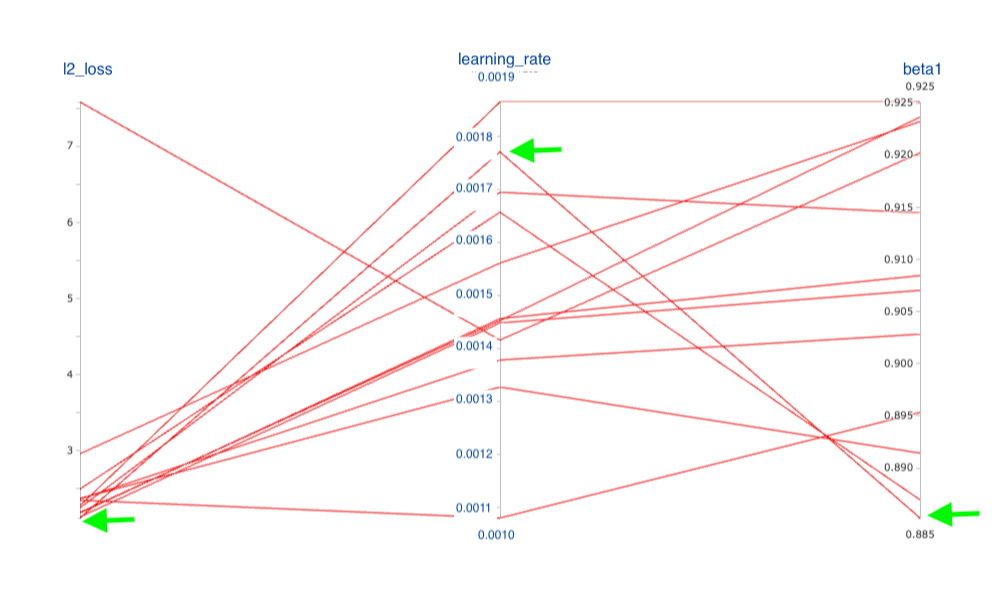}
        \caption{Adam Optimizer in CNN}
        \label{fig:adam_cnn}
    \end{subfigure}
    \begin{subfigure}{0.5\textwidth}
        \centering
        \includegraphics[width=\linewidth]{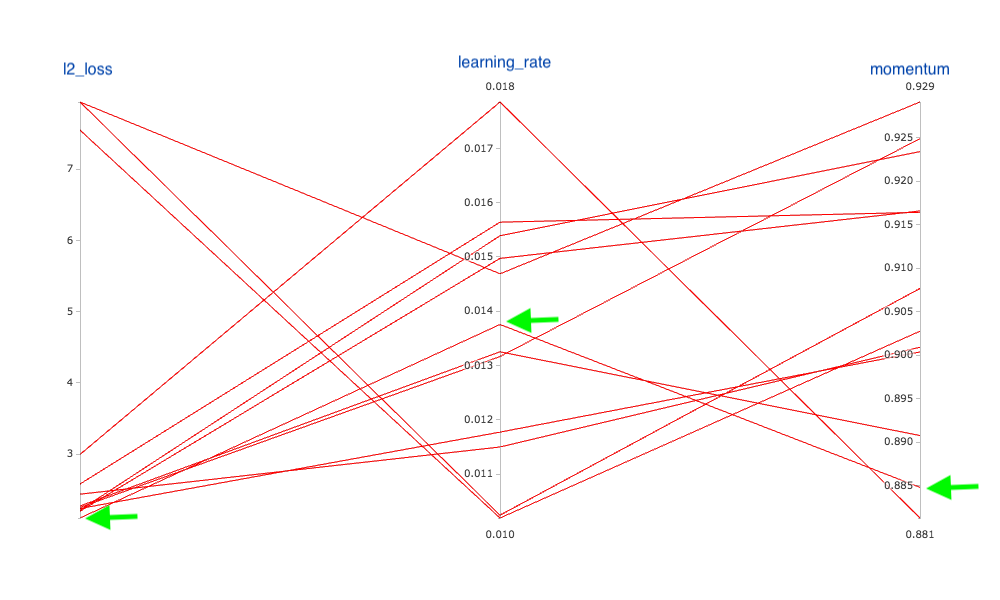}
        \caption{SGD Optmizer in CNN}
        \label{fig:sgd_cnn}
    \end{subfigure}
    \caption{Hyperparameter optimization using Bayesian optimization applied to CNN.
    The arrows mark the optimal values.}
\label{fig:cnn-bayesian-opt}
\end{figure*}

\begin{figure}[t]
    \centering
    \includegraphics[width=\linewidth]{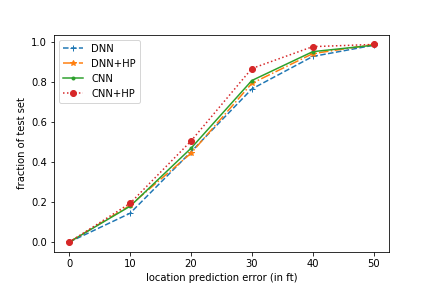}
    \caption{Comparison of CNN and DNN with and without 
    hyperparameter (HP) optimization with the Adam optimizer. A steeper and higher curve is better.}
    \label{fig:dnn_vs_cnn}
\end{figure}

\if 0
\begin{figure*}[t]
\captionsetup[subfigure]{justification=centering}
    \begin{subfigure}{0.5\textwidth}
        \centering
        \includegraphics[width=\linewidth]{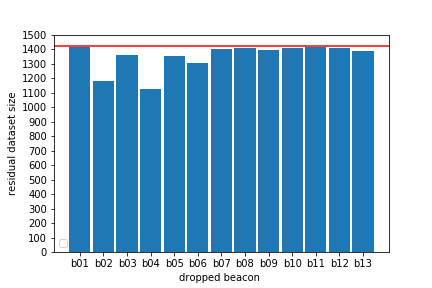}
        \caption{Residual Dataset. Red line shows original data set size of 1420.}
        \label{fig:residual_dataset}
    \end{subfigure}
    \begin{subfigure}{0.5\textwidth}
        \centering
        \includegraphics[width=\linewidth]{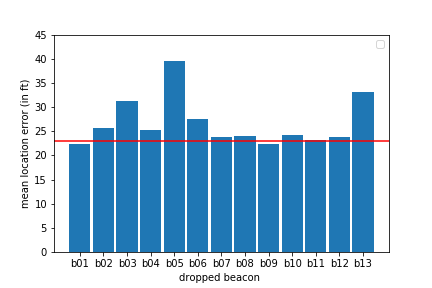}
        \caption{Mean Location Prediction Error. Red line shows error without
        any dropout, 23.2 ft.}
        \label{fig:location_error}
    \end{subfigure}
    \caption{Impact of dropping iBeacons on localization error.}
\label{fig:dropout}
\end{figure*}
\fi 

We took a CNN model available on Kaggle~\cite{mypaper:kaggle-dnn}
and made obvious changes to improve model performance. We changed
the filter sizes, increased the number of feature maps, and replaced last
CNN layer with a dense layer; Figure~\ref{fig:cnn} shows the architecture.
The epochs, batch size, loss function, train:test split, are the same
as specified in Table~\ref{tab:params}. 
In order to apply CNN to the problem of wireless localization, the data is converted
into a gray scale image. Each of the data instance 
(i.e., a row of 13 iBeacons RSSI values corresponding to a specific location)
is transformed into a gray scale image of 
size [25,25,1], corresponding to 25 X 25 grids in which the 
entire floor map (Figure~\ref{fig:setup}) is divided into.
Thus, there as many images as there
are samples in the data set. The image is created via the following steps:
\begin{enumerate}
    \item The image is initialized to a zeroed image of size [25, 25, 1], i.e., an image
    with just black pixels.
    \item Each beacon's sensor location coordinates (x,y,1) is assigned with a pixel value
    that is obtained by dividing the RSSI measurement by -200 (minimum RSSI value). This
    normalizes the pixel values to the [0,1] range.
\end{enumerate}
%The image is initialised to a zeroed
%image of size [25, 25, 1] -- an image with just black pixels. For each sample,
%the normalized RSSI values 
%(1 maps to $-200$ dBm, implying no signal and is represented by white in the image)
%from each of the iBeacons is set at coordinate $<x, y, 1>$, 
%where $x$ and $y$ are the 2-dimensional coordinates of the location of the iBeacon. 
As an example, suppose a receiver located at $<$10, 15$>$ measures an
RSSI of $-20$ dBm for iBeacon $i$, which has a location of $<x_i, y_i>$,
then the value of coordinate $<x_i, y_i, 1>$ is set to the normalized value
of $-20/-200$ dBm. This sample image is then labelled with the $<x,y>$
coordinates of the receiver, $<$10, 15$>$. Such an image is generated for 
each location of the receiver thus making as many images as there are samples in the 
data set.
The CNN is then trained such that given a gray scale image, the neural 
network predicts the corresponding $x$ and $y$ coordinates of the receiver. 

\begin{figure*}[tbp]
\captionsetup[subfigure]{justification=centering}
    \begin{subfigure}{0.5\textwidth}
        \centering
        \includegraphics[width=\linewidth]{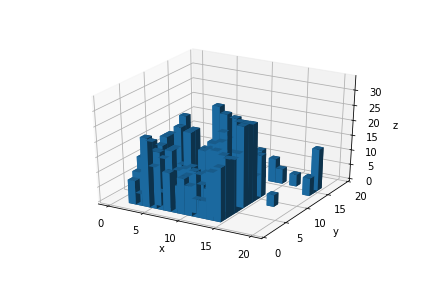}
        \caption{Overall distribution of 1420 samples.}
        \label{fig:dataset_dist}
    \end{subfigure}
    \begin{subfigure}{0.5\textwidth}
        \centering
        \includegraphics[width=\linewidth]{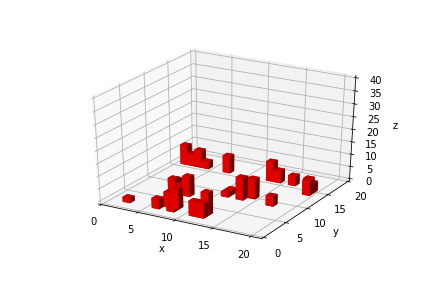}
        \caption{188 under-represented ($<10$ samples) locations in the data set.}
        \label{fig:dataset_underrepresented}
    \end{subfigure}
    \caption{Distribution of labelled samples in the data set.
    The $z$-axis shows the number of samples
        for the corresponding grid (25 X 25) specified by the $x$ and $y$ axes.}
\label{fig:dataset}
\end{figure*}

\begin{figure}[t]
    \centering
    \includegraphics[width=0.9\linewidth]{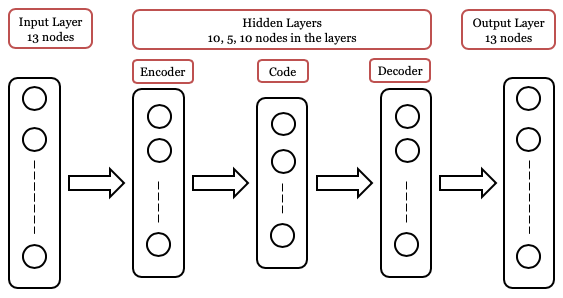}
    \caption{Autoencoder architecture for data augmentation.}
    \label{fig:autoencoder}
\end{figure}

\begin{figure}[t]
    \centering
    \includegraphics[width=\linewidth]{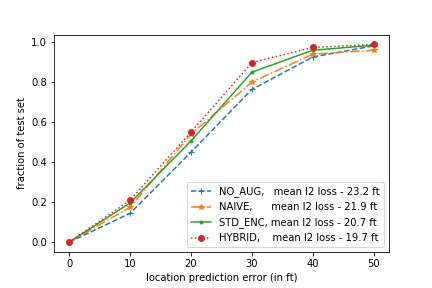}
    \caption{Location Prediction Error with Augmentation for a DNN model with Adam optimizer
    and default hyperparameters. The naive and standard 
    autoencoder based augmentations improve the performance compared to the original
    data set. The hybrid approach shows the most improvement.}
    \label{fig:augmentation}
\end{figure}

\subsection{CNN with Hyperparameter Optimization}
\label{cnn_hp}

Just like in the case of DNN, we apply HP-tuning using
Bayesian optimization to the CNN and the
results are shown in Figure~\ref{fig:cnn-bayesian-opt}. Overall the pattern looks
similar to using hyperparameter optimization with DNN
(Figure~\ref{fig:dnn-bayesian-opt}) but the main difference to note is that 
the range of the metric values (\emph{l2 loss}) is much higher in the case of CNN
than in the case of DNN, with both the Adam and the SGD optimizer. This means
that hyperparameter search is more relevant with the use of CNN than DNN when applied
to the wireless localization problem.

\subsection{CNN and DNN Comparison}
\label{comparison}

Figure~\ref{fig:dnn_vs_cnn} shows the comparison of the different models, 
all with Adam optimizer. The Y-axis
shows the fraction of test samples that were predicted within a prediction error,
shown in the X-axis. All the models perform very similar though there is 
noticeable difference between them. As expected the base DNN model performs the
worst and DNN with HP-tuning
does improve the performance marginally. This is expected 
from row 1 of Table~\ref{tab:hyperparams} that shows that the tuned
\emph{learning rate} and \emph{beta1} are very close
to their respective default values. As the number of beacons increases,
we expect that HP-tuning for DNNs would yield further gains.
The base CNN model shows performance very close to the DNN with HP-tuning but the
CNN with HP-tuning does show a noticeable improvement.

%Given the number of columns/beacons is small in this dataset,
%DNN model seems to be doing better than CNN. 
%If the number of beacons are large, say 100's, DNN approach would result in learning
%large number of parameters. Also, DNN model would need to be remodelled and then
%retrained whenever there is any change in the number of iBeacons.
%CNN models on the other hand, don't need to be remodelled,
%but just need to be retrained. Also, CNNs use less learning parameters.
%Moreover, if there is similar layout (example multiple floors in a building,
%similar placement of beacons and RSSI from different sets of devices)
%transfer learning can be applied.

%It is also observed that CNN model requires a fair amount
%of HP tuning compared to DNN model. {\bf XXX} Any reason why?
%We believe that some more HP tuning can be done on some additional parameters for CNN
%such as number of filters and filter sizes. However, at this time,
%we set them aside for future work.

\emph{The exact DNN or CNN model is not that relevant here.
The most important thing that we
want to convey is that hyperparameter tuning allows a quick and automatic 
search to find the best hyperparameters for any given model.}
%Additionally, as we shall see in the next section, even though a DNN is less generic
%than a CNN due to the DNN's dependence on number of beacons, it is still a very 
%useful model to do quick analysis to figure out which beacons are more useful and 
%which are not.

%% file: tex/augmentation.tex
\section{Dataset Augmentation}
\label{augmentation}
\if 0
\begin{figure*}[htbp]
\captionsetup[subfigure]{justification=centering}
    \begin{subfigure}{0.5\textwidth}
        \centering
        \includegraphics[width=\linewidth]{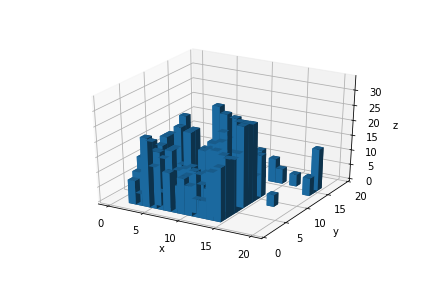}
        \caption{Overall distribution of 1420 samples.}
        \label{fig:dataset_dist}
    \end{subfigure}
    \begin{subfigure}{0.5\textwidth}
        \centering
        \includegraphics[width=\linewidth]{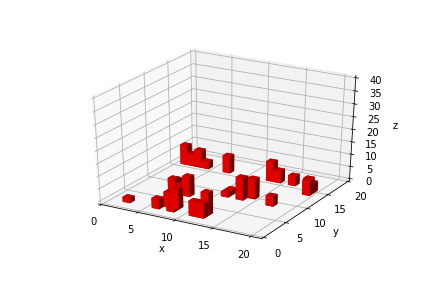}
        \caption{188 under-represented ($<10$ samples) locations in the data set.}
        \label{fig:dataset_underrepresented}
    \end{subfigure}
    \caption{Distribution of labelled samples in the data set.
    The $z$-axis shows the number of samples
        for the corresponding grid (25 X 25) specified by the $x$ and $y$ axes.}
\label{fig:dataset}
\end{figure*}

\begin{figure}[t]
    \centering
    \includegraphics[width=0.9\linewidth]{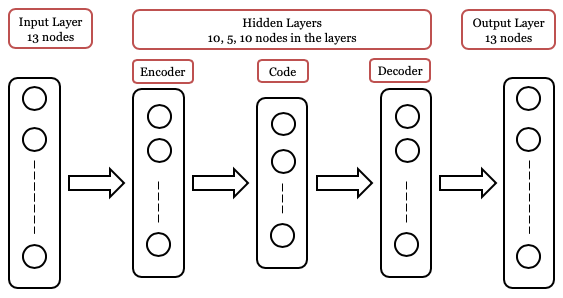}
    \caption{Autoencoder architecture for data augmentation.}
    \label{fig:autoencoder}
\end{figure}

\begin{figure}[t]
    \centering
    \includegraphics[width=\linewidth]{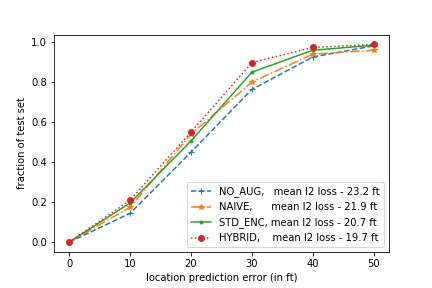}
    \caption{Location Prediction Error with Augmentation for a DNN model with Adam optimizer
    and default hyperparameters. The naive and standard 
    autoencoder based augmentations improve the performance compared to the original
    data set. The hybrid approach shows the most improvement.}
    \label{fig:augmentation}
\end{figure}
\fi
A basic problem with location samples is that compared to 
the total number of samples, only 
a small fraction are labelled. For example, as mentioned earlier in Section~\ref{dataset},
in the BLE data set~\cite{UCI:ble-dataset}, there are only 1420 labelled samples but a 
much larger 5191 unlabelled samples. In general, ML models perform better with larger
training samples.

%In order to augment the data set,
We started by analysing the distribution of samples with respect
to the grid (Figure~\ref{fig:setup}). As shown in Figure~\ref{fig:dataset_dist},
different 
grids have different number of samples and the distribution is not uniform. This is expected
since the library floor has pillars and furniture that makes several grid locations
inaccessible. On top of it since the data collection is done by humans it is simply 
unrealistic to expect equal number of samples at each location. 
%\emph{The target of augmentation is to generate more samples for
%the grid locations with fewer (but not zero) samples,
%as shown in Figure~\ref{fig:dataset_underrepresented}}.
\emph{We need to augment these under-represented 
areas (Figure~\ref{fig:dataset_underrepresented}) for better prediction in these
areas}.

\subsection{Naive Approach}
\label{naive}
We first tried a naive approach to augment the data set. This approach \emph{does not} 
use any unlabelled sample. In this approach, for a location
with fewer samples ($<10$), as shown in Figure~\ref{fig:dataset_underrepresented},
new samples (with the same label, i.e., the same location) are 
generated using a uniform distribution between the minimum and maximum value of the RSSI for
any given beacon. For example, if a location has two samples, one with RSSI values from beacons
\emph{b1, b2,} and \emph{b3}, and the other from just \emph{b1} and \emph{b2}, a new sample
is generated with RSSI values for \emph{b1} and \emph{b2} as a random sample between the
minimum and maximum RSSI values from the two respective samples;
\emph{b3} is skipped since it is not 
present in both the samples and hence the minimum and maximum of the RSSI values
for \emph{b3} are undefined.
We use this to generate \emph{one additional sample} for each of the 188 under-represented
locations.
With this approach we were able to generate another 188 labelled sample points thus giving a 
total of 1420 + 188 = 1608 labelled samples that we used for training and testing. The results
with this augmented data set for a DNN with Adam optimizer and default hyperparameters
are shown in Figure~\ref{fig:augmentation} and the naive approach
shows a distinct advantage over the original data set; mean error reduces from 23.2 ft
to 21.9 ft. However, as already mentioned, this
approach does not utilize the unlabelled samples; we address that issue with the use of 
autoencoders.

\subsection{Standard Autoencoder}
\label{autoencoder}
%Then we tried \emph{autoencoders} to augment the data set.
Autoencoders are a standard 
technique for augmenting data sets and we wanted to check whether it can be used for our data
set.
Goodfellow et al.~\cite{Goodfellow-et-al-2016} define an \emph{autoencoder} as a neural network
that is trained to attempt to copy its input to its output. 
As shown in Figure~\ref{fig:autoencoder}, it has an input layer and an output layer that is 
separated by one ore more hidden layers. The hidden layers constitute
an \emph{encoder}
that encodes the input into a lower dimension \emph{code}
and a \emph{decoder} that tries to reconstruct the original image from the code. A more detailed
description of autoencoders is outside the scope of this paper.

The autoencoder is run for 20 epochs, with each epoch processing the entire set of 5191 
unlabelled images. Thus the autoencoder learns the distribution of the unlabelled data set.
Once this is done, we take one sample for each of the 188 under-represented locations and feed
it to the autoencoder and use the output as an additional sample, with the same label as the
input sample. Though 188 samples were fed, there were some samples generated that were 
showing RSSI values for beacons that the location had not seen in any of the existing samples.
There were 30 such generated samples and these were discarded. 
This left us with 188 - 30 = 158 samples and the labelled data set was augmented
to 1420 + 158 = 1578 samples. Figure~\ref{fig:augmentation} shows that 
the 20.7 ft mean error with this augmented data is slightly better than the 21.9 ft mean
error with the naive approach. 

\label{Hybrid Approach}
\label{hybrid}
The natural thing to do is to combine both the previous approaches into a
\emph{hybrid approach}.
This gives an additional 188 + 158 samples, thus taking the overall count of labelled samples
to 1420 + (188 + 158) = 1766. As expected, and as shown in Figure~\ref{fig:augmentation}, 
this hybrid approach gives the best performance with a mean error of 19.7 ft. Overall,
this amounts to around 15\% reduction in the prediction error. In this work,
we created just one
additional sample for each under-represented location, if needed,
more samples for these locations can be generated.

%% file: tex/conclusion.tex
\section{Conclusion}
\label{conclusion}

In this paper, we have presented two techniques for improving the performance of deep
learning models for wireless localization.
First, we analysed how hyperparameter optimization can be applied to DNN and CNN models to 
optimize their performance and to make them more relevant to wireless environments, which are
naturally very dynamic.
Second, we present two augmentation
techniques to augment the labelled data set so that more data is made available for training 
the deep learning models, thus improving their performance.
Finally, in the interest of open access and further research, we have made all code relevant
to this paper public at
{\it \url{https://github.com/CiscoAI/deep-learning-wireless-localization}}.

%Finally,
%we also showed how we can do analyze which beacons are more important than others that can
%help us in deciding the optimal number of beacons required to achieve practical wireless 
%localization.

%% file: tex/appendix.tex
\subsection{Katib: Hyperparameter Optimization Software}
\label{katib}

\begin{figure}[t]
    \centering
    \includegraphics[width=\linewidth]{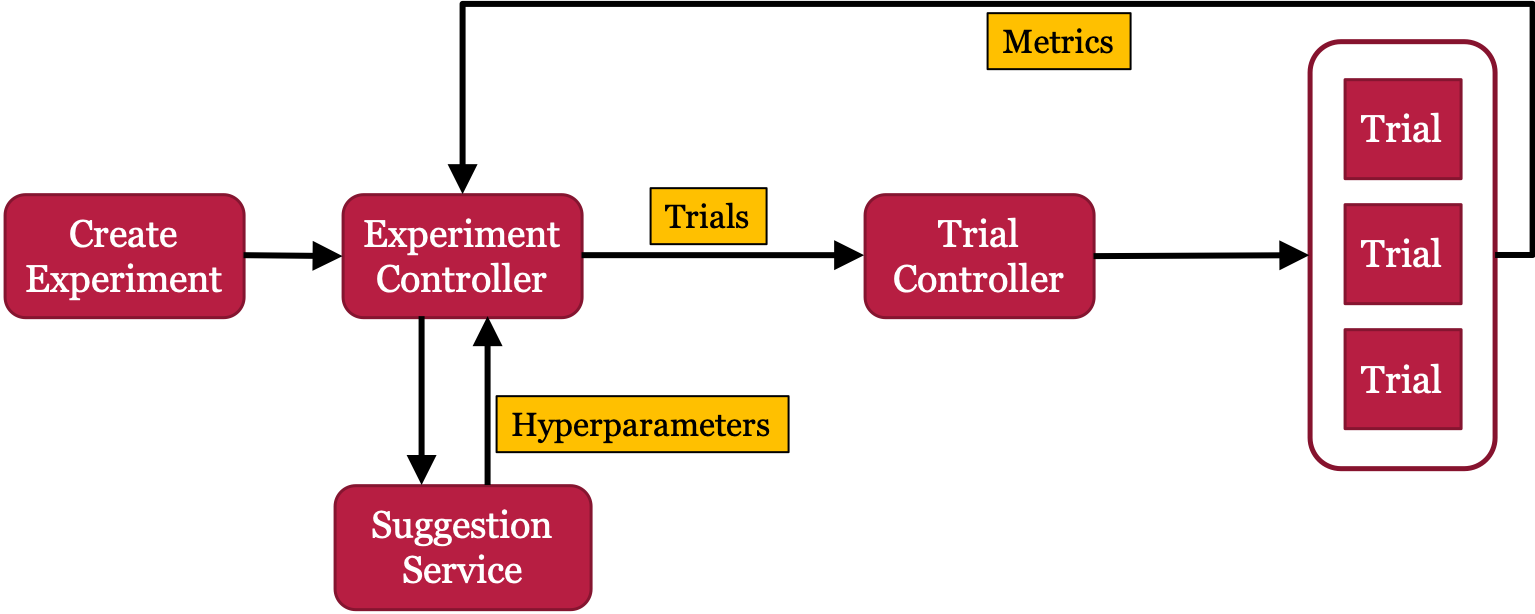}
    \caption{Katib System Architecture.}
    \label{fig:katib}
\end{figure}

We used Katib~\cite{mypaper:katib} as the software for hyperparameter tuning.
It is an open-source Kubernetes~\cite{mypaper:k8s} native system 
for hyperparameter tuning. It is based on Google Vizier~\cite{golovin:vizier-kdd2017} and
is able to support multiple ML frameworks such as Keras~\cite{mypaper:keras},
TensorFlow~\cite{mypaper:tensorflow} and
PyTorch~\cite{mypaper:pytorch}; we used Keras in this paper.
Since Katib is based on Kubernetes, it provides scalability,
portability, and fault tolerance. It also has several hyperparameter tuning algorithms such
as {\it random search, grid search, hyperband, and Bayesian optimization.} Additionally,
the user can plugin any customized algorithm. Moreover, Katib is tightly integrated with
Kubeflow~\cite{mypaper:kubeflow}, an open source ML toolkit for Kubernetes with great
traction from the industry, making Katib (within Kubeflow) an ideal bet for long term 
relevance.

As shown in the schematic of Katib in Figure~\ref{fig:katib},
an experiment is created and a handle to the experiment is sent to the 
{\it Experiment Controller}, which uses a {\it Suggestion Service} to get suggestions
for hyperparameters. These hyperparameters along with the actual ML code form the 
{\it trials}, which are then run by a {\it Trial Controller}. The Trial Controller runs the
trials on the underlying Kubernetes platform, collects the metrics of the run
and feeds the metrics back to the Experiment Controller, which then continues the experiment
with the next set of hyperparameter suggestions from the Suggestion Service. The Suggestion
Service supports standard search methods and can also be fitted with custom search
algorithms.
Even though we use Katib in this paper,
other hyperparameter optimization frameworks, such as Hyperopt~\cite{mypaper:hyperopt},
could be used as well. 

\subsection{Kubeflow: The Machine Learning toolkit for Kubernetes}
\label{kubeflow}

Kubeflow~\cite{mypaper:kubeflow} is an open-source project that 
makes the deployment of ML workflows on any Kubernetes installation simple,
portable, and scalable. It supports running ML jobs written in frameworks such as
TensorFlow~\cite{mypaper:tensorflow} and PyTorch~\cite{mypaper:pytorch} and is
extensible to any other framework. It is one of the most popular open-source
project for managing ML workflows with contributors from a large number of companies.
Even though Katib can be used independently we use Katib on top of
Kubeflow since it eases the experimentation process.